# An Integrated Framework for Sensing Radio Frequency Spectrum Attacks on Medical Delivery Drones

Philip Kulp, Cybrary Fellow, College Park, MD 20740

Nagi Mei, Capitol Technology University, Laurel, MD 20708

ABSTRACT

*Drone susceptibility to jamming or spoofing attacks of GPS, RF, Wi-Fi, and operator signals presents a danger to future medical delivery systems. A detection framework capable of sensing attacks on drones could provide the capability for active responses. The identification of interference attacks has applicability in medical delivery, disaster zone relief, and FAA enforcement against illegal jamming activities. A gap exists in the literature for solo or swarm-based drones to identify radio frequency spectrum attacks. Any non-delivery specific function, such as attack sensing, added to a drone involves a weight increase and additional complexity; therefore, the value must exceed the disadvantages. Medical delivery, high-value cargo, and disaster zone applications could present a value proposition which overcomes the additional costs. The paper examines types of attacks against drones and describes a framework for designing an attack detection system with active response capabilities for improving the reliability of delivery and other medical applications.*

## 1. Introduction

Traditional IT architecture protection consists of specially purposed cybersecurity systems designed for distinct activities such as firewalls and Intrusion Detection System (IDS). Replicating this multi-device protection architecture is impractical in a drone-based system such as medical delivery services, so an embedded attack detection functionality would be required. Solo missions expose the drones to varying attack vectors such as Radio Frequency (RF), Wi-Fi, Global Positioning System (GPS) jamming (Palmer & Geis, 2017), GPS spoofing (He, Chan, & Guizani, 2016), control channel disruption, or other malicious activity (Javaid, Sun, Devabhaktuni, & Alam, 2012). If the control link is unencrypted, a malicious actor could perform a takeover of the drone (He et al., 2016) or leverage a reverse-engineered toolkit to predict the frequency hopping pattern (Kamkar, 2013). The intent of the actor is decoupled from the method of attack in the current research to focus explicitly on identifying and reacting to indicators of malicious activity.

Manufacturers implement multiple safety measures for the drones in case the vehicle loses the connection to the operator, GPS fails, needs to avoid a collision trajectory, or enters restricted airspace (Congress, 2018). The safety features can be exploited as potential attack vectors if a malicious actor forces the sensors into a situation within the parameters that require an emergency protection response. Our research has implications for medical delivery, disaster zones, and the Federal Aviation Administration (FAA) interest in actors committing federal crimes (FCC, 2020; Orleff, 2020). U.S. laws prohibit intentional or malicious interference of authorized radio communications.

The applicable laws are the Communications Act of 1934 (47 USC §333) for radio communications and the Criminal Code for satellite communications. If a person interferes with government communications while jamming a targeted drone, the laws are applicable for fine or imprisonment. A drone is considered an aircraft per FAA Reauthorization Act of 2018 §371 (Congress, 2018); therefore, the Aircraft Sabotage Act (18 USC §32) and Aircraft Privacy (49 USC §§46501-07) apply for the prohibition of attacks against drones.

While a drone may not avoid all instances of malicious attacks, recording telemetry data, and monitoring the stream for signs of attack may provide benefits to multiple parties. The information could equip the owner of the service, the FAA, or law enforcement with data to correlate with other evidence to identify the perpetrator. The sharing of telemetry data among the drones in a fleet could also deliver a stream of information for proactive measures such as area avoidance or active monitoring. If one drone enters into a troubled state (Palmer & Geis, 2017), the other drones in the fleet could avoid the geographic area or enter an active monitoring mode to provide triangulation (Nguyen et al., 2019), video recording, or additional telemetry data recording for later forensic analysis.

## 2. Related works

While some studies have focused on classifying attacks against Unmanned Aerial Vehicles (UAV; Choudhary et al., 2018), our research explored the creation of specific methods for identifying attacks. Studies have confirmed the need for signature-based behavior detection with a formalized method of identifying attacks based on anomalies (Choudhary et al., 2018). Threats to drones require modeling before developing signatures to identify abnormal behavior (Javaid et al., 2012).

The component of the study relied on modeling cybersecurity threats, which focused on our similar topic of identifying methods of attacking the radio frequency components (Javaid et al., 2012). While the research leveraged the perspective of confidentiality, integrity, and availability for the communication link, other researchers provided a narrower study on the attacks against the sensors (Muniraj & Farhood, 2017). During our review, we identified a significant gap in the literature for creating a framework to explicitly identifying communications links on the RF, WiFi, and GPS sensors.

While drones may overcome controller signal tampering, GPS is a critical component that relies on unencrypted communications and can be susceptible to spoofing (He et al., 2016). Our research acknowledges the GPS attacks of spoofing and jamming as critical threats to overcome and dedicated a section to discussing methods of mitigating the assaults. The United States Congress has also identified the susceptibility of GPS as a priority and mandated the need for a resilient service that cannot be corrupted or degraded (National Timing Resilience and Security Act of 2017, 2017). The focus of multiple research studies on GPS vulnerabilities has become paramount based on the declining costs of spoofing, which can be as low as $255 to create an All-GNSS & Map spoofer (US National PNT Advisory Board, 2018).

Malicious actors can also use similar methods of attacking the controller signal via Wi-Fi. The Wi-Fi or other RF control channels present a prime target for the malicious actor since hijacking this communication would allow the attacker to fly the drone (Dressel & Kochenderfer, 2019). Other researchers exploited the communications protocol over the Wi-Fi channel as the attack by identifying three distinct vulnerabilities (Hooper et al., 2016).

The researchers manipulated JavaScript Object Notation (JSON) files sent over UDP to perform Denial of Service (DoS), buffer overflow, and ARP cache poisoning against the drone (Hooper et al., 2016). These types of zero-day attacks may not have an effective defense since the exploits are against the software developed by the drone manufacturer. Since the attack first requires a deauthentication mechanism (Kamkar, 2013) against the valid controller, a signature for detecting this activity in combination with later anomalous activity may be an indicator.

While the motivation of the malicious actor may not be needed to process attacks, some threat vectors such as command link jamming (Palmer & Geis, 2017) may require other reactions. A safety system could engage a lost-link flight path (Palmer & Geis, 2017), which would be the intent of the malicious actor. If the malicious actor knows the contingency plan or alternate path (Zhou & Kwan, 2018), they may force the situation above to exploit a secondary attack. The design of a detection system for drones should include a threat evaluation for each of the attacks and threat vectors.

The application and business function of the drone system require tuning of the sensing system. Research has focused on the innovation of drone delivery in health care (Scott & Scott, 2018), medical delivery in hostile environments (Hooper et al., 2016; Palmer & Geis, 2017), and general delivery systems (Sawadsitnag et al., 2015). With medical delivery systems, actors may be interested in the theft of packages or malicious disruption. In either scenario, the shipment may involve lifesaving needs such as vaccines (Haidari et al., 2016) and would require a sensing system for identifying and averting attacks. Any non-delivery specific function, such as attack sensing, added to a drone involves a weight increase and additional complexity; therefore, the value must exceed the disadvantages. Medical delivery, high-value cargo, and disaster zone application could present a value proposition which overcomes the negative aspects.

## 3. The proposed solution

The solution requires a two-step approach of creating signatures for the detection of attacks and defining a framework for performing active measures against the attacks. The signatures would be pair with a graduated scale of active measures based on the severity and assurance an attack is occurring. The framework describes the data requirements, storage dependencies, hardware, and software to implement the solution. The solution includes the topic of a drone swarm, which could provide additional information when a malicious actor attacks a drone in the fleet. Finally, the solution addresses the need for a graduated scale of response levels based on the levels of detected attacks, which would precipitate actions for the drone under attack or others within the system.

The drone attack sensing system is composed of several components in an orchestrated pipeline, which requires input from multiple sources with additional processing units, as depicted in the following figure.

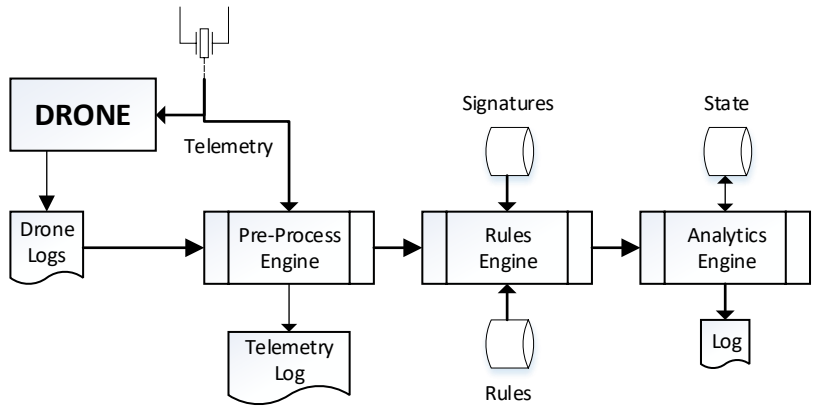

**Fig. 1.** Architecture of the attack sensing and processing pipeline

The *pre-process engine* is responsible for ingesting raw telemetry data from antennas to normalize the disparate data sources into a common format. The pre-processor could also parse the manufacturer's onboard logs (Clark, Meffert, Baggili, & Breitinger, 2017). The *pre-process engine* stores the raw data into logs for future analysis and forwards the normalized data to the *rules engine*. The *rules engine* uses the signatures to tag events for processing and correlation to the *analytics engine*. Finally, the *analytics engine* maintains stateful information about events to determine the need for actions or countermeasures.

*3.1. Telemetry data*

Enabling detection capabilities of a system would require the collection of data in the form of events used for streaming analysis or future review. The detection system would not collect data from all possible frequencies since the interesting events would correlate with the existing RF sensors that represent an attack vector. The interesting events in this context refer to sensors on the drone, which could represent attack vectors such as Wi-Fi, RF control, or GPS signals. Consumer First Person View (FPV) drones use a radio frequency in the Wi-Fi band of 2.4 GHz or 5.8 GHz to communicate and transmit video back to the controller (Di Pietro, Oligeri, & Tedeschi, 2019). Alternate frequencies used for communication are 900 MHz and 1.3 GHz (Di Pietro et al., 2019). Additionally, some drones use the Sik telemetry radio, which implements Frequency Hopping Spread Spectrum (FHSS) on the 915 MHz ISM band (Dressel & Kochenderfer, 2019).

Custom software could generate the telemetry information matched with custom hardware or extracted from existing data generated by the manufacturer of the drone. The DROne Parser (DROP) opensource project provides functionality to read the DJI Phantom III drone logs (Clark et al., 2017). While the primary goal of the project is to provide a forensic capability for parsing the logs, near real-time telemetry data may be available from the existing drone sensors, which could provide data events for signature analysis. The DROP project is an example for the specific drone manufacturer, but other forensic projects are available, which could provide data for additional manufacturers (Mantas & Patsakis, 2019).

*3.2. Pre-process engine*

The pre-process engine would receive telemetry data from multiple sources and normalize the data before simultaneously forwarding to the rules engine and writing to a log file. Table 1 contains the recommended event and telemetry data based on the minimum requirements for drone attack signature matching. Each log entry would allocate a variable width of text for the data, separated by whitespace, similar to the format of web server logs ("Log Files - Apache HTTP Server," n.d.).

**Table 1. Telemetry log composition**

| Data type | Format |
| --- | --- |
| Timestamp | 2020-03-01T19:40:08Z (ISO, 2019) |
| Speed | The current speed of drone in km/hr |
| Direction | The direction of travel in m/s |
| Geo | Latitude, longitude, altitude (ISO, 2008) |
| Selector | Taxonomy of selectors |
| Additional | Additional context for selector data (e.g., frequency selector could use the field for dB) |

The log event data contains a 'selector' entry, which would leverage a pre-defined taxonomy of terms used to group events. The log includes a 'selector data' field, which contains expanded information from the 'selector' field (e.g., 'frequency' selector would consist of the MHz/GHz, and the 'additional' field would include the power measured). The 'selector' field would improve the attack detection analysis engine performance by reducing the need to execute regex or another text-based matching on non-applicable events. Table 2 contains a sample list of selectors for grouping log events.

**Table 2. Log selectors**

| Selector |
|---|
| DEBUG |
| EMERGENCY |
| FREQUENCY |
| GENERAL |
| SIGNAL_LOSS |

*3.3. Rules engine*

In the proposed system, every drone would collect telemetry data used for monitoring attacks, or in the case of an active attack, drones may collect additional telemetry. Similar to a traditional IDS, signatures detect attacks based on known indicators or unusual activity. Each signature would be composed of two components, the indicator and level of action required. Table 3 contains a recommended hierarchy of actions based on four levels of monitoring.

**Table 3. Attack action labels**

| Level | Action |
|---|---|
| Info | Data used for later analysis |
| Elevated | Additional logging initiated |
| Group | Trigger nearby drones to enter elevated |
| Emergency | Emergency procedures with mandatory notification to nearby drones or mothership |

The second component of the signature includes the indicator to identify potentially malicious activity in the event log data. The rules engine would continuously compare the stream of events against the signature database and use the indicators to match data, which triggers an alert for processing by the *analytics engine*. Table 4 contains an abbreviated list of sample signatures for possible drone attacks (Akos, 2012).

**Table 4. Drone IDS sample signatures**

| Level | Signature |
|---|---|
| Info | Lost link (GPS, WiFi, RF) |
| Elevated | WiFi frequencies > X power at a distance Y from controller |
| Elevated | Deauthenticaion of the WiFi (Nguyen, Nguyen, Tran, Vu, & Mittal, 2008) |
| Group | DDoS UAV network detection (Miquel, Condomines, Chemali, & Larrieu, 2017) |
| Emergency | GPS DoS/Spoofing: 1575.42 MHz signal strength > -120dB (Akos, 2012; Gaspar, Ferreira, Sebastiao, & Souto, 2018) |

Example signatures include the determination a GPS spoofer may be active using several techniques with varying requirements, complexity, hardware, or software. The strength of an incoming signal, which is greater than the normal noise floor of -120 dB, may indicate malicious activity (Akos, 2012). Measuring the angle of arrival with multiple antennas ("Security of Global Navigation Satellite Systems," n.d.) could indicate a spoofed signal. An indicator includes the number of visible satellites, which increases during a spoofing attack above a usual threshold of 4 to 8 (Warner & Johnston, 2003). The spoofer needs to overpower the real signals, so they can send false information from up to 10 fictitious satellites which would be another indication of false information (Warner & Johnston, 2003).

*3.4. Analytics engine*

The *analytics engine* would perform advanced processing of the signatures which fired during the *rules engine* processing. Some signatures may only require a single event to trigger an alert that requires action, but other signatures may provide indicators that require monitoring. For example, in the case of Wi-Fi monitoring, an anomalous signal may be indicated by unusual power at a specific altitude or distance from the operator. The *analytics engine* would then store the state of this event and create an alert if the signature fired a second time after a distance, as indicated by the time passed and speed of the drone.

The state database would serve multiple functions within the monitoring system. As described above, a database would maintain the state of signatures for identifying attacks that require multiple time-based events. The second storage of the state would maintain a sliding window of telemetry metadata. The metadata storage creates a concise but minimized record of data similar to netflow (Yang et al., 2020). Netflow is a metadata collection of network packets traversing the network without full storage of the actual content. While metadata collection may significantly reduce data storage requirements, it could still overwhelm the limited storage capacity of the drone.

For the proposed system, we recommend a sliding window (Papapetrou, Garofalakis, & Deligiannakis, 2012) of metadata collection, which maintains fresh data of a limited period with continuous overwriting. The *analytics engine* would access the telemetry metadata data based on rules matching events but would not need access to stale data. New incoming data would overwrite the stale data to reduce data storage requirements (Papapetrou et al., 2012) after a period, such as 1 hour.

*3.5. Hardware requirements*

A unidirectional antenna would provide triangulation capabilities for the drones with an associated software pattern to control the movements of the drone. A single antenna and the rotational maneuvering capabilities would allow the drone swarm to provide an approximate triangulation of rogue RF signals (Nguyen et al., 2019). A single rotating antenna can detect a spoofed signal using a spatial processing method of anisotropy to decipher authentic or spoofed signals (Wang, Li, & Lu, 2017). Limitations to the triangulation of RF derive from the surrounding terrain and buildings, which can cause issues such as multipath problems (Wright, Freedman, & Liu, 2010).

An alternate method of triangulation would use Time Difference of Arrival (TDoA) to triangulate a signal (FCC, 2020). TDoA would require accurate synchronization among devices such as GPS or signals from a central location. Even in the case of GPS unavailability because of jamming, TDoA can utilize GPS holdover for a short period of minutes while still maintaining accuracy (Podevijn et al., 2018). Another alternative to multiple antennas would be the use of Software Defined Radio (SDR; (Di, Morton, & Vinande, 2013), discussed in the software section.

*3.6. Software requirements*

While custom software may be required to orchestrate the translation from telemetry data to log collection, multiple open-source projects or research efforts exist, which could facilitate the other major components such as the *rules engine* and *analytic engine*. The *analytic engine* could be based on knowledge graphs and leverage the design of vlog (Carral et al., 2019). Research has also focused on improving the performance in pattern matching for rule engines (Zhao & Bai, 2016).

Kismet is a Wi-Fi mapping tool ("Kismet," n.d.) that can leverage the OpenWrt project and take advantage of the customizable framework (Raviglione, Malinverno, & Casetti, 2019). Researchers have used the OpenWrt framework as the backbone of experiments for testing different vehicle communications systems (Abunei, Comsa, & Bogdan, 2016). The various research projects could be combined and tailored to adapt the findings to drone communications.

Leveraging SDR (Schmidt, Akopian, & Pack, 2018) with GNU Radio (Orleff, 2020) could assist in the development of a radio capable of receiving different frequency bands (Marwanto, Sarijari, Fisal, Yusof, & Rashid, 2009). With improvements to embedded systems, the technological requirements for signal processing in an SDR are achievable. universal software radio peripheral (USRP), which is a commercial hardware platform for SDR (Marwanto et al., 2009), could provide additional improvements in the capabilities.

USRP performs the analog to digital conversion of the radio waves (Di et al., 2013)and passes the signal to the GNU Radio software for processing. GNU radio then performs filtering using the signal processing blocks to demodulate the signal into the original data (Marwanto et al., 2009). Software processing capabilities could evaluate the GPS signals to monitor satellite signal strength or extract the identification codes (Warner & Johnston, 2003). GPS signal processing would be valuable for identifying attacks and also switching to alternate data sources as a countermeasure to an attack.

*3.7. Monitoring modes*

Table 5 includes the designated modes a single or swarm allocated drone could enter, which would be triggered by an attack event.

**Table 5. Drone Operation sensing modes**

| Mode |
| --- |
| Normal |
| Monitor |
| Elevated |
| Evasive |
| Swarm Monitor |
| Swarm Elevated |

Normal mode represents the standard operating method in which the drone operates according to the business function. The monitor mode indicates a slight elevation in the alertness of the drone, which requires additional capture of data for monitoring or collection. A signature triggered by the detection of an attack causes an elevated mode, and the drone begins operating in an active defense model, which could impact the business function. Distress represents the most elevated mode of operation. A forced landing due to mechanical malfunction, controller channel attack, or other situations could trigger an alert based on high data usage such as a Distributed Denial of

Service (DDoS) attack on the network (Miquel et al., 2017). The high data requirements are necessary to capture audio, video, or photos of the immediate scene to obtain evidence of the malicious activity.

As previously indicated, a swarm mode would allow a distressed drone to request assistance, which would trigger other nearby drones to enter a slave mode and receive event logs. The event logs could include telemetry data useful for activities such as RF triangulation (Nguyen et al., 2019) or audio/visual data capture to assist in evidence collection. The swarm mode could mimic the single drone usage modes of monitor and elevated.

The mode would also require an exit stage whereby the drones could reduce the monitoring level. The monitoring level reduction would be required to avoid maintaining a higher level and exhausting storage, disrupting business activities, or any other negative consequences. The de-escalation of operating modes would be for single and swarm mode.

*3.8. Countermeasures*

Similar to the event horizon of a black hole whereby even light cannot escape, logs may need to be continuously shared into an audit repository using log forwarders to the centralized database (Adike & Vamshi Krishna, 2019) to avoid data lost once successful jamming begins. The telemetry data from a continuously jammed drone may never be retrieved if the recorded data does not leave the drone, and the attacker destroys the evidence. When sensing a potential attack, a drone could enter an emergency communications mode in which it falls back to alternate methods of transmitting data such as 3G, SMS, or WiMAX (Ali, Fatima, & Biradar, 2018). With additional components, the drone could shift to the use of an Automatic Gain Controller (AGC), which varies the gain of the amplifier to isolate weaker signals and avoid saturation (Akos, 2012). The process may not succeed, but the AGC technique could be used in an emergency to attempt to lock onto a valid GPS signal for the temporary purpose of evading the attack.

A drone could overcome GPS jamming by switching to a different guidance system such as the European Galileo, Chinese Bei Dou, Indian Regional Navigation Satellite System (IRNSS), or Russian Global Navigation Satellite System (GLONASS; Palmer & Geis, 2017). The use of SDR and USPR, when used in combination with GNU Radio, could leverage a single antenna capable of receiving multiple GPS systems of global navigation satellite systems (GNSS; Di et al., 2013). A system with the capability to utilize multiple satellite-based navigation systems (Di Fonzo, Leonardi, Galati, Madonna, & Sfarzo, 2014) could switch to alternate source if the primary became unavailable and the signature-based detection identified the failure as a jamming or spoofing attack.

Researchers have achieved acceptable gain when testing a fractal slot antenna across a wide range of frequencies from 1.47 GHz to 10.73 GHz (Marwanto et al., 2009). The frequencies are within the ranges of the GPS L1 signal (1.57542 GHz) and the GLONASS L1 signal (1.602 GHz). The researchers did not explicitly test frequency bands within the range of the L2 satellite signals of GPS (1.2276 GHz), GLONASS (1.246 GHz), or other GNSS. Further experimentation in the area could test the ability to extend an SDR system into the wider band.

The drone could also collect audio/visual data in several different monitoring modes. A default time-lapse capture mode would minimize data storage requirements while implementing automated processing (Shakeri & Zhang, 2017). Once the drone entered a heightened state, based on an attack signature trigger, elevated monitoring would occur. Photo, video, or audio data of the scene could be captured or streamed to the nearby drones or the mothership to protect the information in case the drone is lost. During a sensed attack, the drone or all geographically close drones could go into monitoring mode to collect evidence. Or if the drone was forced into a landing situation, for the perpetrator to steal the cargo, audio/visual transmission of the scene could provide evidence for law enforcement.

**4. Conclusions**

Our research identified valuable methods of detecting attacks for single events (Akos, 2012; Nguyen et al., 2008; Palmer & Geis, 2017); trend identification is also essential. Satellites exhibit distinguishable trends unique from spoofing equipment such as the inclusion of identification codes, satellite signal strength, and time intervals (Warner & Johnston, 2003). The integration of the *analytics engines* in the proposed solution takes these factors into account by providing trend analysis to identify attacks that may not exhibit anomalous behavior in a single instance. For example, a GPS spoofer would create an artificially constant time interval, whereas a valid satellite signal is subject to atmospheric interference, which would not exhibit this behavior (Warner & Johnston, 2003).

Overcoming the value proposition would require benefits above the features defined in the current study. The theft of over 90,000 packages occurs in New York City on any given day, and the delivery services absorb the losses as part of doing business (Hu & Haag, 2019). The additional cost of software, hardware, research, and engineering to enable the attack detection should not be cost-prohibitive. The cost versus value proposition may also be overcome by delivery systems that carry high-value packages or other sensitive material. Medical deliveries (Haidari et al., 2016), especially in rural or disaster areas, could represent packages of a time-sensitive nature that overcome the cost/benefit analysis of the delivery service providers. Drone delivery service providers could realize cost saving by coordinating services (Sawadsitnag et al., 2015) and notifications of malicious activity among each other or with the FAA.

The applicability of the research extends to multiple disciplines or implementations of drone systems, including, medical delivery, or military applications of health delivery in hostile environments. The implementer would tune the signature database to the specific use, based on threat modeling to identify the types of expected attacks. Thresholds could be tuned in the *rules engine* or *analytic engine* to reduce false-positive rates. Metropolitan areas would require tuning to account for terrain or other features where large scale and differing power levels of Wi-Fi signals (Nguyen et al., 2008) may produce false positives.

Future research should include designing studies to evaluate signatures not tested in the referenced research. While individual signature testing could produce valuable data, focusing on the *analytic engine* processing may build the most significant knowledge. Configuration and tuning of events over time could reduce the false positive rate and improve detection capabilities. Finally, performing future studies to evaluate the benefits of implementing sensing technology in drone swarm systems could produce additional areas of study. Research on drone swarm studies could focus on correlating data from multiple sources into a single analytics engine. The individual drones could also improve the overall reaction capabilities by coordinating monitoring capabilities.